\documentclass{article}

\def\fnote#1#2{\begingroup\def\thefootnote{#1}\footnote{#2}
    \addtocounter{footnote}{-1}\endgroup}
\def\sato{mmsato11@yahoo.co.jp}

\usepackage[pdftex]{graphicx}
\usepackage{amsmath}
\usepackage{amsfonts}

\begin{document}

\pagestyle{empty}
\vspace{16pt}
\begin{center}
{\Large \bf Renormalization Group Transformation for Hamiltonian Dynamical Systems in Biological Networks}\\
\vspace{16pt}


Masamichi Sato\fnote{*}{\sato}
\vspace{16pt}

{\sl Dai Nippon Printing, Co., Ltd.}
\vspace{12pt}\\
{\bf ABSTRACT}
\vspace{12pt}\\

\begin{minipage}{4.8in}
We apply the renormalization group theory to the dynamical systems with the simplest example of basic biological motifs.
This includes the interpretation of complex networks as the perturbation to simple network.
This is the first step to build our original framework to infer the properties of biological networks, and the basis work to see its effectiveness to actual complex systems.
\end{minipage}

\end{center}
\vfill
\pagebreak

\pagestyle{plain}
\setcounter{page}{1}

\baselineskip=16pt

\section{Introduction}

In ref.~\cite{mochizuki}, Mochizuki, et al. established a criterion to extract informative nodes in biological networks.
Our work has been inspired by their work, and we consider the application of physical theory to reduce a complex network to simpler network.
Unlike the work of ref.~\cite{mochizuki}, our framework does not propose any discipline to reduce networks to a simpler one, but our work is used to analyze the transformation after a reduction, at least.

Dynamical systems is a common tool in many fields of sciences, not only biology, but also other fields which needs control theory.
This system is described by ODEs and many researchers have been studying its phenomenological such as chaos or complex systems, and also its theoretical aspects such as graph theory.
The concerns of current study are its theoretical aspects.
We study the properties of dynamical systems using the simplest motifs which appear in biological systems.
The reason to analyze the simplest motifs is we should start to study the fundamental properties which are inferred by study with our original way.
Our method of study is based on physical theory.
Especially, we show the results obtained from inferences by renormalization group theory and the implications to theoretical biological study brought by using them.
Though, our analysis is tend to concentrate on the simplest case to much, but this is because we should start to study on basic properties brought by adopting our methods, as aforementioned, then we should apply current method to an actual systems.
However, this remains as our future work.

This paper is organized as follows:
In section two, we give a description of perturbative dynamical systems and its renormalization.
Section three is devoted for the description of RG transformation to perturbed fundamental motifs.
Section four is the coherent potential-like picture which appears in perturbation theory of disordered physical systems.
Section five on symmetry breaking.
In section six, we give several comments.
Section seven is conclusions.

\section{Perturbative dynamical systems and renormalization}

We start with the definition of dynamical systems.
Dynamical system is defined as a set of ODEs.
Linear case is described as

\begin{equation}
\dot{{\bf x}}=H {\bf x}.
\end{equation}

\noindent
Here, ${\bf x}$ is the vector of varying variable and $H$ is a matrix called as Hamiltonian~\cite{hamiltonian}, which takes constant elements describing the changes of variables.
The matrix $H$ is called as reaction coefficients matrix for chemical reaction systems.
Non-linear case is described as 

\begin{equation}
\dot{{\bf x}}=Hf({\bf x}).
\end{equation}

Here, we introduce a perturbation to this system.
we consider the perturbation taking the form of additional term to Hamiltonian.

\begin{equation}
\dot{{\bf x}}=(H+F)f({\bf x})
\label{pert}
\end{equation}

If the perturbation $F$ is small enough to be described as the form proportional to $H$, that is,  

\begin{equation}
F=\epsilon H
\end{equation}

\noindent
the eq.~(\ref{pert}) becomes

\begin{eqnarray}
\dot{{\bf x}}&=&(1+\epsilon) Hf({\bf x})\\
&=&R\cdot Hf({\bf x}).
\end{eqnarray}

\noindent
For this case, the difference of the perturbed elements of Hamiltonian from the original is the proportional factor and this is interpreted as just the modification to coupling constant or scale.
This case is renormalizable and we call $R$ as ``renormalization transformation''~\cite{RG}.

\section{RG to perturbed fundamental motifs}

Next, we consider the several concrete examples of renomalization.

\subsection{Basic reaction}

We begin with the simplest case.

\begin{figure}[h]
  \centering
	\mbox{
	\includegraphics[width=4cm, bb=0 0 175 84]{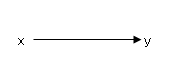}
	}
	\caption{Basic reaction}
	\label{basic_arrow}
\end{figure}

If we consider the reaction depicted as fig.~\ref{basic_arrow}, this reaction is described by the following equation.

\begin{equation}
\dot{{\bf x}}=K {\bf x}
\end{equation}

\noindent
Here, ${\bf x}$ is the vector of

\begin{equation}
{\bf x}=(x, y)^T
\end{equation}

\noindent
and the marix $K$ is

\begin{equation}
K=\left (
\begin{array}{cc}
-k_1&0\\
k_1&0
\end{array}
\right ).
\label{stoichi_k}
\end{equation}

\noindent
The constant elements of matrix $K$ are called reaction coefficients. 

\begin{figure}[h]
  \centering
	\mbox{
	\includegraphics[width=4cm, bb=0 0 174 102]{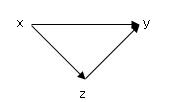}
	}
	\caption{Basic loop reaction}
	\label{basic_tri}
\end{figure}

Next, we consider a little bit more complex case.
This is depicted as fig.~\ref{basic_tri}.

\noindent
This reaction system is described by the ODEs of

\begin{equation}
\dot{\tilde{{\bf x}}}=\tilde{K} \tilde{\bf x},
\end{equation}

\noindent
with

\begin{equation}
\tilde{\bf x}=(x,y,z)^T,
\end{equation}

\noindent
and

\begin{equation}
\tilde{K}=\left (
\begin{array}{ccc}
-k_1-k_2&0&0\\
k_1&0&k_3\\
k_2&0&-k_3
\end{array}
\right ).
\end{equation}

\noindent
Here, we introduce the three-dimensional extension of eq.~(\ref{stoichi_k}) such as 

\begin{equation}
K'=\left (
\begin{array}{ccc}
-k_1&0&0\\
k_1&0&0\\
0&0&0
\end{array}
\right ).
\end{equation}

\noindent
$K'$ is related with $\tilde{K}$ by the following transformation:

\begin{equation}
K'=\tilde{R}\cdot \tilde{K}.
\end{equation}

\noindent
A little bit algebra realizes the concrete form of $R$ as follows,

\begin{equation}
\tilde{R}=\left (
\begin{array}{ccc}
r_{11}&((r_{11}-1)k_1+r_{11}k_2)/(k_1+k_2)&((r_{11}-1)k_1+r_{11}k_2)/(k_1+k_2)\\ 
r_{21}&((r_{21}+1)k_1+r_{21}k_2)/(k_1+k_2)&((r_{21}+1)k_1+r_{21}k_2)/(k_1+k_2)\\ 
r_{31}&r_{31}&r_{31}
\end{array}
\right ).
\end{equation}

\noindent
This shows that $\tilde{R}$ can be interpreted as the renormalization transformation described in the last section.
In fact, $\tilde{R}$ is decomposed into the addition of perturbative term to original $K'$ for small $k_2$ and $k_3$.  
The added dimension of $z$ can be interpreted as perturbation to the original 2-dimensional reaction.

\subsection{1-loop}

\begin{figure}[h]
  \centering
	\mbox{
	\includegraphics[width=10cm, bb=0 0 481 301]{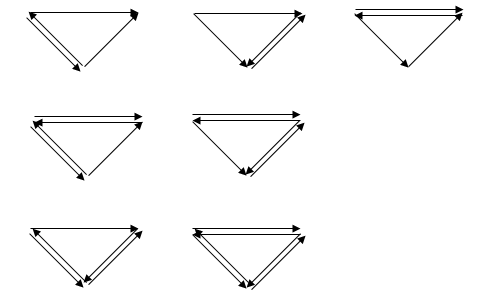}
	}
	\caption{1-loop motifs}
	\label{1_loop_m}
\end{figure}

Here, we list all of reactions which consist of forward- and reverse-actions with 1-loop of fundamental motifs~\cite{alon}.
These are depicted in fig.~\ref{1_loop_m}.

The reaction coefficients matrices corresponding the above figures are listed below.

\begin{eqnarray}
\left (
\begin{array}{ccc}
-k_{12}-k_{13}&0&k_{31}\\
k_{12}&0&k_{32}\\
k_{13}&0&-k_{31}-k_{32}
\end{array}
\right )
\;
\left (
\begin{array}{ccc}
-k_{12}-k_{13}&0&0\\
k_{12}&-k_{23}&k_{32}\\
k_{13}&k_{23}&-k_{32}
\end{array}
\right )
\;
\left (
\begin{array}{ccc}
-k_{12}-k_{13}&k_{21}&0\\
k_{12}&-k_{21}&k_{32}\\
k_{13}&0&-k_{32}
\end{array}
\right )
\nonumber
\end{eqnarray}

\begin{eqnarray}
\left (
\begin{array}{ccc}
-k_{12}-k_{13}&k_{21}&k_{31}\\
k_{12}&-k_{21}&k_{32}\\
k_{13}&0&-k_{31}-k_{32}
\end{array}
\right )
\;
\left (
\begin{array}{ccc}
-k_{12}-k_{13}&k_{21}&0\\
k_{12}&-k_{21}-k_{23}&k_{32}\\
k_{13}&k_{23}&-k_{32}
\end{array}
\right )
\;
\begin{array}{ccc}
{}{}{}\\
{}{}{}\\
{}{}{}\\
\end{array}
\nonumber
\end{eqnarray}

\begin{eqnarray}
\left (
\begin{array}{ccc}
-k_{12}-k_{13}&k_{21}&k_{31}\\
k_{12}&-k_{21}-k_{23}&k_{32}\\
k_{13}&k_{23}&-k_{31}-k_{32}
\end{array}
\right )
\;
\left (
\begin{array}{ccc}
-k_{12}-k_{13}&k_{21}&k_{31}\\
k_{12}&-k_{23}-k_{21}&k_{32}\\
k_{13}&k_{23}&-k_{31}-k_{32}
\end{array}
\right )
\;
\begin{array}{ccc}
{}{}{}\\
{}{}{}\\
{}{}{}\\
\end{array}
\;
\begin{array}{ccc}
{}{}{}\\
{}{}{}\\
{}{}{}\\
\end{array}
\;
\begin{array}{ccc}
{}{}{}\\
{}{}{}\\
{}{}{}\\
\end{array}
\;
\begin{array}{ccc}
{}{}{}\\
{}{}{}\\
{}{}{}\\
\end{array}
\;
\begin{array}{ccc}
{}{}{}\\
{}{}{}\\
{}{}{}\\
\end{array}
\;
\begin{array}{ccc}
{}{}{}\\
{}{}{}\\
{}{}{}\\
\end{array}
\;
\begin{array}{ccc}
{}{}{}\\
{}{}{}\\
{}{}{}\\
\end{array}
\;
\begin{array}{ccc}
{}{}{}\\
{}{}{}\\
{}{}{}\\
\end{array}
\;
\begin{array}{ccc}
{}{}{}\\
{}{}{}\\
{}{}{}\\
\end{array}
\nonumber
\end{eqnarray}

For these reactions, the description of renormalization transformation also holds and it's easy to calculate the concrete form of them.
However it's space costing, so we do not show them here.

Another aspect that we should notice is the structure of symmetry breaking.
Adding an arrow to the basic diagram of fig.~\ref{basic_tri} corresponds to the addition of one parameter.
This is representing the structure causing symmetry breaking.

\subsection{2-loop}

\begin{figure}[h]
  \centering
	\mbox{
	\includegraphics[width=3.5cm, bb=0 0 138 136]{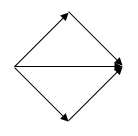}
	}
	\caption{2-loop motif}
	\label{2_loop_m}
\end{figure}

The fig.~\ref{2_loop_m} is an example of 2-loop reaction.
The reaction coefficients matrix for this reactions is:

\begin{eqnarray}
\left (
\begin{array}{cccc}
-k_{12}-k_{13}-k_{14}&0&0&0\\
k_{12}&0&k_{32}&k_{42}\\
k_{13}&0&-k_{32}&0\\
k_{14}&0&0&-k_{42}
\end{array}
\right ).
\end{eqnarray}

This case is also renormalizable and the addded three- and four-dimension can be interpreted as a perturbation to the original two-dimensional reaction system.

\section{The coherent potential-like picture}

We consider the reaction which is constituted with the all of one-loop perturbations.
Taking the form of Hamiltonian as,

\begin{equation}
H=K+F,
\end{equation}

\noindent
$K$ is the original two-dimensional reaction and $F$ corresponds to the perturbation.
$K$ is represented in three-dimensional form as,

\begin{eqnarray}
K=\left (
\begin{array}{ccc}
-k_{12}&0&0\\
k_{21}&0&0\\
0&0&0\\
\end{array}
\right ),
\end{eqnarray}

\noindent
and $F$ is represented with the elements of $K_{ij}$

\begin{eqnarray}
F=\left (
\begin{array}{ccc}
K_{11}&K_{12}&K_{13}\\
K_{21}&K_{22}&K_{23}\\
K_{31}&K_{32}&K_{33}\\
\end{array}
\right ).
\end{eqnarray}

\noindent
$K_{ij}$ are represented by the elements summed over all of 1-loop matrices~\cite{yonezawa}.

\begin{eqnarray}
K_{11}&=&-k^1_{13}-k^2_{13}-k^3_{13}-k^4_{13}-k^5_{13}-k^6_{13}-k^7_{13}\\
K_{12}&=&k^3_{21}+k^4_{21}+k^5_{21}+k^7_{21}\\
K_{13}&=&k^1_{31}+k^4_{31}+k^6_{31}+k^7_{31}\\
K_{21}&=&0\\
K_{22}&=&-k^2_{23}-k^3_{21}-k^4_{21}-k^5_{21}-k^5_{23}-k^6_{21}-k^6_{23}-k^7_{23}-k^7_{21}\\
K_{23}&=&k^1_{32}+k^2_{32}+k^3_{32}+k^4_{32}+k^5_{32}+k^6_{32}+k^7_{32}\\
K_{31}&=&k^1_{13}+k^2_{13}+k^3_{13}+k^6_{13}+k^7_{13}\\
K_{32}&=&k^2_{23}+k^5_{23}+k^6_{23}+k^7_{23}\\
K_{33}&=&-k^1_{31}-k^1_{32}-k^2_{32}-k^3_{32}-k^4_{31}-k^4_{32}-k^5_{32}\nonumber \\
&&-k^6_{31}-k^6_{32}-k^7_{31}-k^7_{32}
\end{eqnarray}

\noindent
By the transformation $\tilde{R}$, $K+F$ is transformed into some matrix $\tilde{K}$ of the same form as $K$,

\begin{equation}
\tilde{R}(K+F)=\tilde{K}
\end{equation}

\noindent
This shows that $K+F$ is renormalizable.
The constituting elements $k_{ij}$ in $K_{ij}$ is reflecting the restrictions by the arrows in diagrams.
So, the perturbations cannot take arbitrary form and is allowed under the considering diagrams, also this is reflecting the all of the contributions and their reacting form of considering perturbative reactions.

\section{Symmetry breaking}

\begin{figure}[h]
  \centering
	\mbox{
	\includegraphics[width=3.5cm, bb=0 0 134 66]{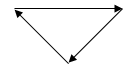}
	}
	\caption{Symmetric motif}
	\label{sym}
\end{figure}

We consider on symmetry breaking for perturbation.
The most symmetric motif for triangle is fig.~\ref{sym}.

\noindent
The corresponding matrix to this diagram is as follows:

\begin{eqnarray}
\left (
\begin{array}{ccc}
-k_{12}&0&k_{31}\\
k_{12}&-k_{23}&0\\
0&k_{23}&-k_{31}
\end{array}
\right ).
\end{eqnarray}

\begin{figure}[h]
  \centering
	\mbox{
	\includegraphics[width=3.5cm, bb=0 0 131 65]{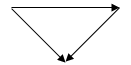}
	}
	\caption{Symmetry broken motif}
	\label{non-sym}
\end{figure}

\noindent
Then we consider the least symmetry broken case.
One of this case is represented as fig.~\ref{non-sym}.

\noindent
The corresponding matrix is such as:

\begin{eqnarray}
\left (
\begin{array}{ccc}
-k_{12}-k_{13}&0&0\\
k_{12}&-k_{23}&0\\
k_{13}&k_{23}&0
\end{array}
\right ).
\end{eqnarray}

What we can infer from this simplest example is the following simple claim: the symmetry breaking of one degree of freedom reduces the rank by one.
So, what we can refer to the analysis of metabolic pathway from this claim is that the metabolic pathway described with general  diagram accompanies the reduction of rank of corresponding matrix caused by how the symmetry is broken.
There might hide a biological reason in the diagram of broken symmetry in the way of symmetry breaking.
The expressing mechanisms appearing in networks of actual biological systems might be able to be inferred by the classification of types of symmetry breaking realized in actual biological systems.

\section{Comments}

\subsection{Renormalization group transformation}

The renormalization transformations shown in above  can be interpreted as coarse graining process and they are inreversible.
In fact, it is easy to see

\begin{equation}
\det R=0,
\end{equation}

\noindent
so the inverse transformation does not exist.
Therefore these form a semi-group.

In higher dimension, we can consider sequential transformations.
If we take the order $1\rightarrow 2$ which follows from the higher to lower dimension, the sequential transformation of 

\begin{equation}
R_2\cdot R_1
\end{equation}

\noindent
is executable, but

\begin{equation}
R_1\cdot R_2
\end{equation}

\noindent
is not executable.

\subsection{Variational RG transformation to general Hamiltonian dynamical systems}

Ref.~\cite{ising_rbm} is treating Ising models and they have shown the relationship between RBM (restricted Boltzmann machine) and variational RG transformation (RG transformation with parameter changes to realize maximum likelihood estimates of parameters of true distribution).
If we adopt this transformation to Hamiltonian matrix instead of image data in their framework, it will realize a more general transformation to Hamiltonian matrix.

\begin{equation}
H \xrightarrow[R_{RG}]{}\tilde{H}
\end{equation}

\begin{equation}
R_{RG}=T_{RBM}
\end{equation}

\subsection{Inference to informative nodes}

In the reference of Mochizuki et al.~\cite{mochizuki}, they are proposing a method of coarse graining of biological reaction networks.
A simple description of their proposal is the reduction of network diagram which includes cyclic graphs to that consists of only acyclic graphs.
This corresponds to some special cases of our renormalization transformation.
The objective of their method is to grab the graphical network with biological meaning.
Our method can be applied after determining the objective network structures.
In this sense, our proposing method is weak.
If we admit their proposal to reduce to the biologically meaningful networks, our method is effective to obtain the concrete description of transformation.

\section{Conclusions}
In this paper, we gave a description of renormalization group analysis of dynamical systems and its implications to biological systems.
As mentioned in introduction, the current study is limited to see its properties on the simplest motif.
To see its effectiveness to actual systems, we have to apply our formalism to actually existing networks in biological systems.
This remains as our future work.

\begin{center}
    {\Large {\bf Acknowledgements} }\\
\end{center}
We greatly thank to kind hospitality of my colleagues for giving ideas and comfortable environments.
\vspace{.1in}

\pagebreak

\end{document}